\documentclass[letterpaper]{emulateapj}
\usepackage{apjfonts}
\usepackage{lscape} 
\usepackage{amsmath,amsfonts}
\usepackage{graphicx}
\usepackage[dvips]{color}
\usepackage[breaklinks=true]{hyperref}
\usepackage{natbib} 
\bibpunct[,]{(}{)}{;}{a}{}{,} 

\newcommand{\msol}{\ensuremath{{\rm M}_{\odot}}}
\newcommand{\Msun}{\msol}

\newcommand{\NB}{{\sc NBODY4}}

\shorttitle{Gravitational waves from binary IMBHs}
\shortauthors{Amaro-Seoane et al}

\begin{document}

\title{Gravitational waves from eccentric intermediate-mass black hole binaries}

\author{Pau Amaro-Seoane\altaffilmark{1},
M. Coleman Miller\altaffilmark{2}, 
Marc Freitag\altaffilmark{3}}

\email{Pau.Amaro-Seoane@aei.mpg.de, miller@astro.umd.edu, freitag@ast.cam.ac.uk} 

\altaffiltext{1}
{(PAS) Institut de Ci{\`e}ncies de l'Espai, IEEC/CSIC, Campus UAB, 
Torre C-5, parells, $2^{\rm na}$ planta, ES-08193, Bellaterra, Barcelona and
Max Planck Institut f\"ur Gravitationsphysik
(Albert-Einstein-Institut), Am M{\"u}hlenberg 1, D-14476 Potsdam, Germany 
}
\altaffiltext{2}
{(MCM) Department of Astronomy and Maryland Astronomy Center for
Theory and Computation, University of Maryland, 
College Park, MD 20742-2421, USA}
\altaffiltext{3}
{(MF) Institute of Astronomy, University of Cambridge, Madingley
Road, CB3~0HA Cambridge, UK}

\label{firstpage}

\begin{abstract}

If binary intermediate-mass black holes (IMBHs; with masses between 100 and
$10^4 \Msun$) form in dense stellar clusters, their inspiral will be detectable
with the planned Laser Interferometer Space Antenna (LISA) out to several Gpc.
Here we present a study of the dynamical evolution of such binaries using a
combination of direct $N$-body techniques (when the binaries are well
separated) and three-body relativistic scattering experiments (when the
binaries are tight enough that interactions with stars occur one at a time). We
find that for reasonable IMBH masses there is only a mild effect on the
structure of the surrounding cluster even though the binary binding energy can
exceed the binding energy of the cluster. We demonstrate that, contrary to
standard assumptions, the eccentricity in the LISA band can be in {\em some} cases as
large as $\sim 0.2 - 0.3$ and that it induces a measurable phase difference from circular
binaries in the last year before merger. We also show that, even though energy
input from the binary decreases the density of the core and slows down
interactions, the total time to coalescence is short enough (typically less
than a hundred million years) that such mergers will be unique snapshots of
clustered star formation.

\end{abstract}



\section{Introduction}

The existence of intermediate-mass black holes (IMBHs; masses $M\sim
10^{2-4}~M_\odot$) is not as certain as that of stellar-mass or supermassive
black holes because there is as yet no conclusively established dynamical
mass for any candidate, although there is strong circumstantial evidence for
this mass range in several cases (see \citealt{MillerColbert04} and references
therein for a review). Mergers of IMBHs would, however, be strong sources of
gravitational waves.

The best studied scenario is the runaway growth of a star in a young
cluster via physical collisions among the most massive stars in the center,
which have sunk through mass segregation
\citep{PortegiesZwartMcMillan00,GurkanEtAl04,PortegiesZwartEtAl04,
FreitagEtAl06}. Recently, \cite{GFR06} addressed the same configuration but
added a fraction of primordial binaries to the stellar system.  Using a
Monte-Carlo stellar-dynamics code, they found that not one but two very massive
stars grow in rich clusters in which 10\% or more of stars are in primordial
hard binaries, suggesting the formation of two IMBHs. However, this result
has not been confirmed yet using more accurate direct $N-$body simulations.
\cite{PortegiesZwartEtAl04} have a simulation with primordial binaries but they
do not see this formation, though it is also currently unclear how different
core concentrations will affect binary IMBH formation with a certain fraction
of primordial binaries.  It is also possible that wind losses may drive
away mass more rapidly than it accretes through further collisions
(see \citealt{Belkus07}), although this relies on uncertain extrapolations
from the $\sim 120~M_\odot$ that is the top of their range (see
their Table~2) to the
$\sim 2000~M_\odot$ masses observed in N-body simulations.

\cite{FregeauEtAl06} considered for the first time the possibility that such a
binary could be observed thanks to the emission of gravitational waves in the
coalescence phase and estimated that one can expect the {\it Laser
Interferometer Space Antenna} (LISA) to detect tens of
them depending on the distribution of cluster masses and densities.
\cite{ASF06} addressed the evolution of a binary of two IMBHs formed as the
result of the collision of two independent stellar clusters and followed the
parameters of the binary orbit down to the region in which it will emit
gravitational waves in the $\sim 10^{-4}-10^{-1}$~Hz 
LISA domain. To do this, they combined
direct-summation simulations with an analytical model to evolve the binary from
a point in which it was hard. 

Here we assume that an IMBH binary has been produced in a
single dense stellar cluster, and study the subsequent sinking of the IMBHs and
the evolution and properties of the binary when it forms. In
\S~2 we discuss our numerical method, which combines direct $N$-body studies
with three-body scattering integrations.  In \S~3 we discuss the astrophysical
implications of our results.

\section{Evolution of the IMBH pair: Numerical method}

\subsection{Direct $N-$body simulations}

Direct $N-$body codes integrate all gravitational accelerations in a
stellar system without supposing any special symmetries.  They are
thus the most general and robust tools for numerical analysis of
stellar clusters \citep{Aarseth99,Aarseth03}.  The code we use,
{\NB}, includes a variety of sophisticated approaches that improve
speed and accuracy, including KS regularization \citep{KS65}, as
well as {\em triple} (3-body subsystems), {\em quad}
(4-body subsystems), and {\em chain regularization}
\citep{Aarseth99,Aarseth03}. It also does not make use of any
softening, which would lead to unrealistic evolution of the orbital
parameters of the binary of massive black holes. The disadvantage of
this or any direct $N-$body code is the required computational
time. However, our calculations are accelerated thanks to the
special-purpose hardware GRAPE-6A single PCI cards of the AEI
cluster {\sc Tuffstein} used for the simulations.  Each card has a
peak performance of 130 Gflops \citep{GRAPE6A}, so that a single
node is comparable to a cluster of $\sim$100 individual CPUs working in
parallel.

\begin{table}
\begin{center}
\begin{tabular}{|l||*{8}{c|}c}
\hline
 Model & ${\cal N}_{\star}$ & ${\cal M}_{\rm bin}/M_{\odot}$ & $\rho_0\,(M_{\odot}/{\rm pc}^3)$  
&  $W_{0}$ & IMF & ${a_0}$ (pc) 
\\ \hline\hline
${A}$  & 30002  & 300+300  & $6\cdot 10^4$    & 6  & single & 0.01  \\
\hline
${B}$  & 30002  & 1000+1000 & $2.4\cdot 10^3$  & 6  & single & 0.01  \\
\hline
${C}$  & 30002  & 300+300  & $2.6\cdot 10^3$  & 6  & Kroupa & 0.1  \\
\hline
${D}$  & 30002  & 300+300  & $6\cdot 10^4$    & 6  & single & 0.1  \\
\hline
${E}$  & 128002 & 300+300        & $10^5$    & 6  & single & 0.3  \\
\hline
${F}$  & 128002 & 1900 + 380 & $10^5$    & 6  & Kroupa & 0.1  \\
\hline
\end{tabular}
\end{center}
\caption{
Initial conditions for our featured direct-summation $N-$body models.  
${\cal N}_{\star}$ is the number of stellar particles used. The mass
of the binary, normalized to solar masses, is given in the third column,
$\rho_0$ is the initial mass density at a distance of 0.1 pc, $W_{0}$ is the
King parameter \citep{King66}. All cases are
single-mass but for model ${C}$ and $F$, in which we have a mass
function, specifically a 5-Myrs evolved Kroupa IMF of masses 0.2, 0.5, 50 and
exponents 1.3 and 2.3 \citep{Kroupa00b}. The 7th column shows the initial semi-major
axis of the binary in pc.
\label{tab.NbodyModels}}
\end{table}

Table~\ref{tab.NbodyModels} gives the initial conditions for the
different simulations that we feature.  We ran six cases with
varying number densities and concentrations, of which two had a
\cite{Kroupa00b} mass function instead of single-mass stars.  The
IMBHs have equal mass except in simulation $F$, which has a mass
ratio of 5.  In our simulations the individual time steps led to
fractional energy errors that were always less than $10^{-4}$ 
per N-body unit of time and,
globally, the total energy error of the cluster (i.e. the
accumulated error in the integration of {\em all} particles) is
$0.015\%$ in the case of our fiducial model ($A$).

\subsection{Evolution of the binary: gravitational radiation versus dynamics}

\begin{figure}
\resizebox{\hsize}{!}{\includegraphics[clip]
{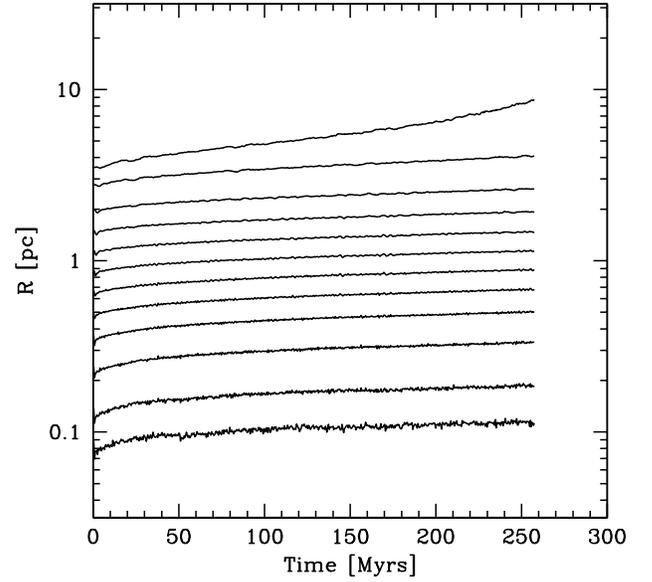}}
\caption{
Lagrangian radii showing the evolution of different mass fractions in
the cluster for Model~D. The fractions are, from the bottom to the top,
$0.01\%$, $0.03\%$, $0.1\%$, $0.2\%$, $0.3\%$ \ldots $0.9\%$ and $0.95\%$
\label{fig.LagRad}}
\end{figure}

Our approach is to evolve the cluster up to $\sim 50-70$ Myrs using the direct
$\NB$ code with a central bound binary. As we discuss below, we can see in
Fig.(\ref{fig.LagRad}) that the stellar cluster experiences a very
moderate expansion during the hardening of the IMBHs.  Once the IMBHs are hard
enough relative to each other they can be treated as an isolated binary that
interacts occasionally with a passing star.  We note that although the
subsequent evolution of the binary will not be identical to that from
the N-body runs, due to the stochastic nature of the encounters, the
general development is similar.  Such interactions tend to increase
the binding energy of the binary, hence shrinking its semi-major axis.  The
eccentricity is also changed, both by Newtonian three-body interactions (which
can increase or decrease the eccentricity; see \citealt{SesanaEtAl07} for a
recent treatment) and by gravitational radiation, which circularizes the
binary.  The combination of the two determines the eccentricity of the binary
when it enters the frequency range of LISA.

\begin{figure}
\resizebox{\hsize}{!}{\includegraphics[clip]
{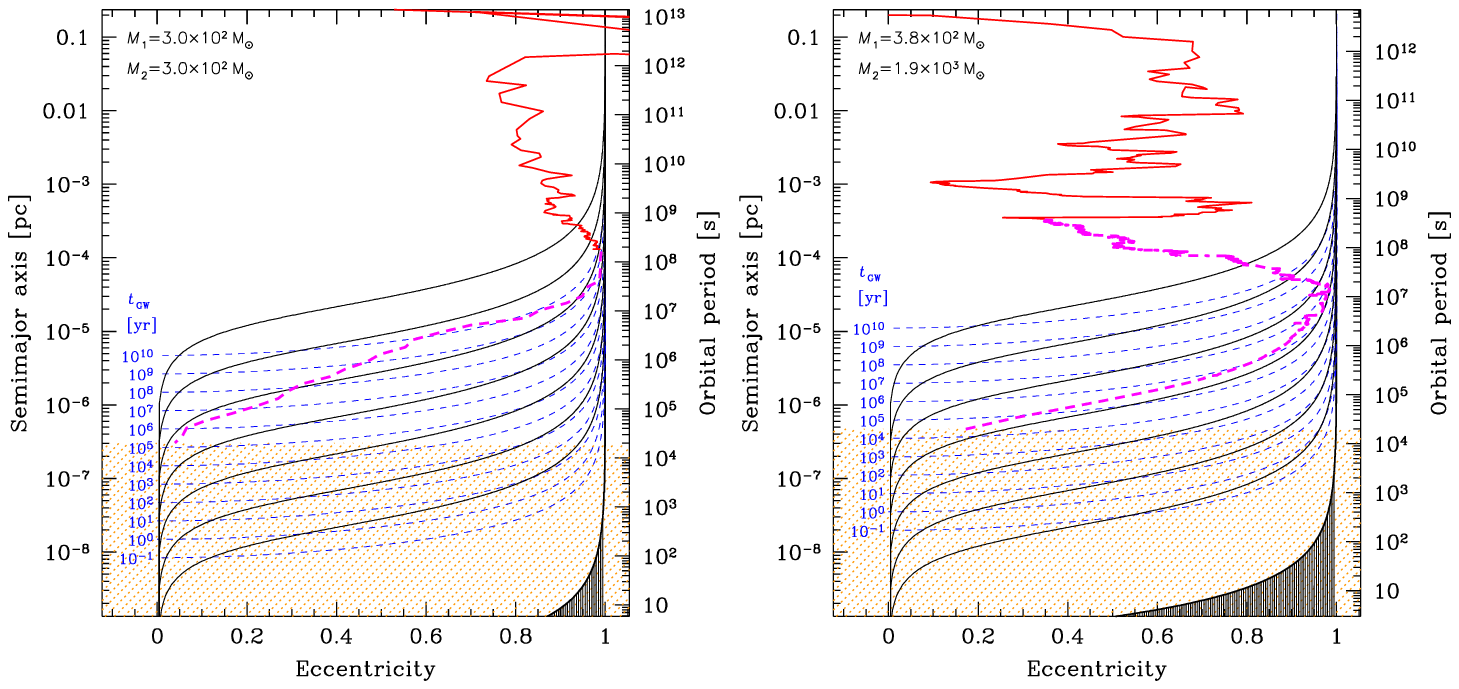}}
\caption{
{\em Left panel:} Inspiral of the IMBH binary of Model~E followed in
the eccentricity--semi-major axis plane. The irregular line shows the
results of the $N-$body simulation. The smooth black solid curves are
the estimated trajectories due to gravitational wave emission
following the approximation of \cite{Peters64}, and the
dashed curves show the corresponding inspiral timescale, $t_{\rm GW}$. 
The dark dashed area depicts the region of unstable
orbits. The lightly shaded area corresponds to the phase in the
evolution in which the $n=2$ harmonic of the gravitational wave signal 
is in the LISA band. The dashed irregular line starting after the
last point of the results of the $N-$body simulation 
(in the color version depicted in magenta), are the 
results from the scattering experiments. See text for further details.
Near the beginning the eccentricity temporarily exceeds unity
because the black holes are not yet bound to each other.
{\em Right panel:} Same for Model~F, which is one
case in which we have initially a Kroupa IMF (see Table\,{\ref{tab.NbodyModels}}).
When star with a big mass interacts with the binary, the eccentricity change is
substantial. The eccentricity therefore wanders up and down, and when it
becomes large enough the binary has a greater chance to spiral together by
gravitational radiation
\label{fig.FiducialAndModelC_A_and_E}
}
\end{figure}

In Figure~\ref{fig.FiducialAndModelC_A_and_E} we show the inspiral of the binary
for models A and C. The irregular lines correspond to the $N-$body simulations.
We take the last point of these evolutions and the number density of field
stars as input to relativistic scattering experiments which we performed
following \cite{GMH06}.  The equations of motion we use for the three-body
encounters include relativistic precession to first post-Newtonian order, as
well as radiation reaction caused by gravitational waves.  Between encounters,
we evolve the semimajor axis and eccentricity of the IMBH binary using the
Peters quadrupolar formulae \citep{Peters64}.  The stars that interact with the
binary are sent with a velocity at infinity of $v$=10~km~s$^{-1}$, typical of
cluster velocity dispersions.  The interaction time is drawn from an exponential
distribution with a mean time of $\tau=(n\Sigma v)^{-1}$, where $n$ is the
stellar number density (taken from the $N-$body simulations) and $\Sigma$ is
the scattering cross section including gravitational focusing. The typical
region in which the IMBH binary wanders is larger than its radius of influence,
hence there is no loss cone as there is for supermassive black holes.

\begin{figure}
\resizebox{\hsize}{!}{\includegraphics[scale=1,clip]
{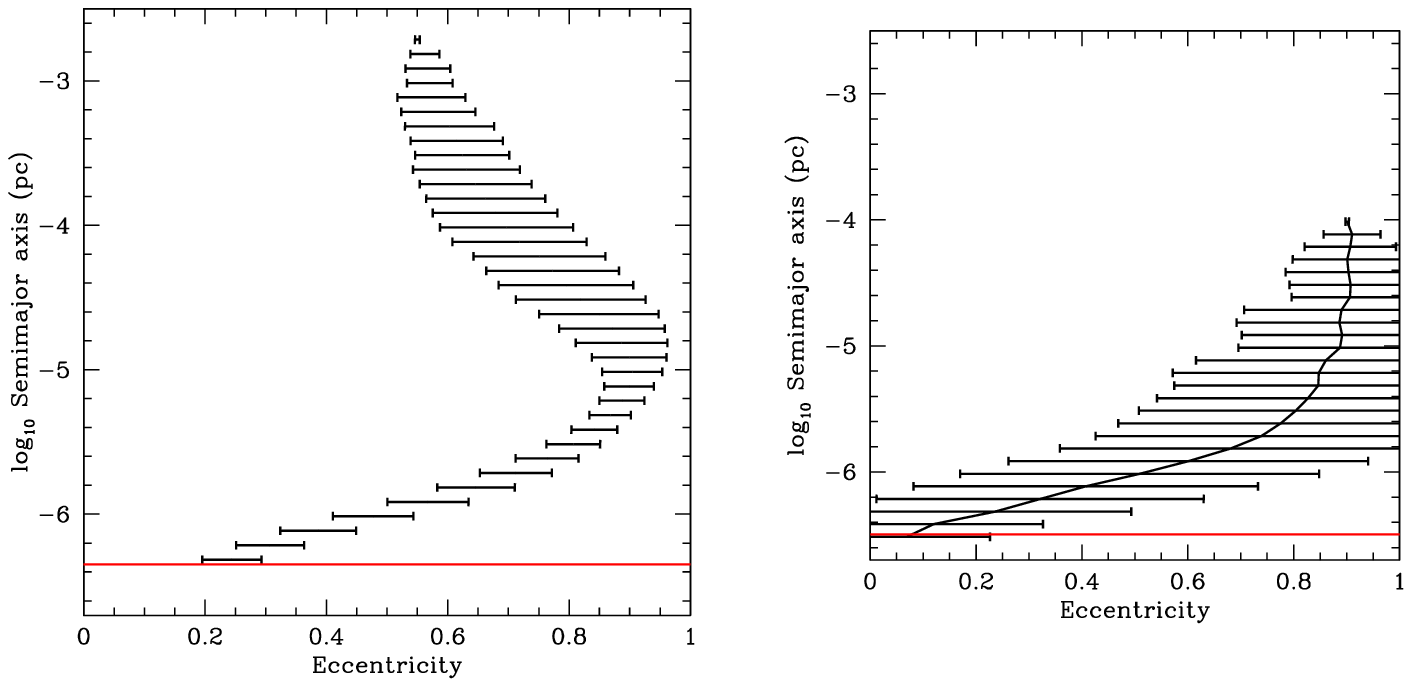}}
\caption{
{\em Left panel:} Average evolution of eccentricity as a function of semimajor
axis. This figure shows the results of 40 three-body scattering experiments,
starting with IMBH masses of $10^3~M_\odot$ each at an initial semimajor axis
of 400~AU and an initial eccentricity of 0.55, corresponding to the end point
of Model~B.  We see the $-1\sigma$ to $+1\sigma$ ranges of eccentricity.  The
horizontal line shows a semimajor axis of 0.093~AU, which is where the orbital
frequency is $5\times 10^{-5}$~Hz and thus the dominant gravitational wave
frequency is 1e-4 Hz.} {\em Right panel:} Similar but assuming a Kroupa IMF and
using the end point of Model~C.  There were no ejections of the binary from the
cluster, which was assumed to have an escape speed of 50 km/s. The horizontal
line is lower than in the left panel because the binary mass is less.
\label{fig.aeavg}
\end{figure}

We ran four sets of 40 simulations, two sets starting at large
separations with zero eccentricity that established agreement with
the direct-summation $N-$body simulations and two sets at the
endpoints of Models~A and B.  In none of these runs
was the binary itself ejected from the cluster, as expected given
its large mass.  For the Model~A endpoint run (binary
mass $600~M_\odot$, initial semimajor axis $a_0=10$~AU, and initial
eccentricity $e_0=0.6$), the eccentricity was $e_{\rm LISA}=0.30\pm
0.10$ when the gravitational wave frequency (equal to twice the
orbital frequency) was at the $10^{-4}$~Hz low end of the {\it
LISA} band; for the Model~B endpoint run (binary mass
$2000~M_\odot$, initial semimajor axis $a_0=400$~AU, and initial
eccentricity $e_0=0.55$) we found $e_{\rm LISA}=0.24\pm 0.05$.  We
show the envelope of the Model~B endpoint runs in the left panel of
Figure~\ref{fig.aeavg}.  We also did scattering experiments
corresponding to the endpoint of Model~C, which had a Kroupa mass
function.  The results are shown in the right panel of 
Figure~\ref{fig.aeavg}.  Compared to the single-mass runs we
see considerably greater variance in the eccentricity as a function
of semimajor axis, and although the range of eccentricities in 
the {\it LISA} band $e_{\rm LISA}=0.11\pm 0.11$ overlaps those in
the single-mass runs there are also a number of cases in which
the binary is nearly circular by the time the gravitational
wave frequency reaches $10^{-4}$~Hz.  This could be a general
feature of scattering interactions when there is a broad mass
function, but we have not performed enough runs to determine
this with confidence.

\section{Discussion and Conclusions}

We have addressed the inspiral of two massive black
holes in a single young stellar cluster.  Our three main results
are: (1)~the cluster itself experiences only mild structural
changes as a result of the inspiral, (2)~the coalescence takes
a short enough time (typically $<$100~Myrs) that mergers occurs
close to the time of formation of the cluster, and (3)~there is
a significant residual eccentricity by the time the binary enters
the LISA band.  We now discuss these conclusions in order.

The stability of clusters against IMBH mergers is consistent with analytic
expectations even though the binding energy of the IMBH binary can exceed the
total binding energy of the cluster by a significant factor. To see this,
consider a circular IMBH binary of component masses $m_1$ and $m_2\leq m_1$,
with total mass $M_{12}=m_1+m_2$ and reduced mass $\mu=m_1m_2/M_{12}$.  As
shown by \cite{Quinlan96}, a star with low speed at infinity that interacts
with the binary will typically be ejected with a speed $v_{\rm ej}\approx
0.85\sqrt{m_2/M_{12}}V_{\rm orb}$, where $V_{\rm orb}$ is the relative speed of
the two objects.  For equal masses $m_1=m_2$, this is $v_{\rm ej}\approx
0.6V_{\rm bin}$.  Suppose now that the cluster has an escape speed $v_{\rm
esc}$.  If $v_{\rm ej}<v_{\rm esc}$ then the star will be retained and share
its kinetic energy with the cluster.  Otherwise, the star will be ejected from
the cluster without depositing significant energy, because the dynamical time
of escape is much less than the relaxation time (which is the time required for
the star to give up energy).  The binding energy of the IMBH binary when
$v_{\rm ej}= v_{\rm esc}$ will be $E_{\rm bin}={1\over 2}\mu V_{\rm
orb}^2\approx {1\over 2}\mu (v_{\rm esc}/0.85)^2(M_{12}/m_2)\approx 0.7 m_1
v_{\rm esc}^2$.  In comparison, if the cluster has a three-dimensional velocity
dispersion $\sigma_{\rm 3D}$ and a mass $M_{\rm cl}$, the binding energy of the
cluster is $E_{\rm cl}\approx {1\over 2}M_{\rm cl}\sigma_{\rm 3D}^2$.  The
ratio is then $E_{\rm bin}/E_{\rm cl}\approx \left(m_1/M_{\rm
cl}\right)\left(v_{\rm esc}/\sigma_{\rm 3D}\right)^2$.  Typically $v_{\rm
esc}\sim 2-3\times \sigma_{\rm 3D}$, so only if the larger black hole mass is
$m_1>0.1-0.2M_{\rm cl}$ could the release of energy unbind the cluster.  We also
note that subsequent to this point, the loss of mass from stars being thrown
out would also soften the cluster.  However, since typically interaction with
of order the binary mass changes the semimajor axis by a factor of $\sim 2$,
just $\sim 10~M_{12}$ in stars will shrink the binary by enough of a factor to
produce coalescence.  Therefore, as verified by our numerical simulations,
hardening of an IMBH binary has only a minor effect on the cluster.

For the time to merger, we note that hardening from large separations
to a few hundred AU takes $\sim 50$~Myr, based on our simulations.
Our three-body runs then indicate that the total time from that point
to merger is virtually always less than 10~Myr, meaning that conservatively
the total time from formation to merger is less than $10^8$~yr.
This is significantly shorter than the age of the universe.  One
consequence of this is that if star formation in massive clusters
was more common at redshift $z\sim 1$ than it is now, and if binary
IMBH formation was also thus more common, then LISA observations of
IMBH mergers will serve as a unique snapshot of star formation as well
as of cluster dynamics (see also \citealt{FregeauEtAl06}).

Figure~(\ref{fig.aeavg}) shows that the eccentricity of the binary will be in
the range $\sim 0.1-0.3$ when the dominant gravitational wave frequency is of
$10^{-4}$ Hz.  Consistent with \cite{Quinlan96} we find that the eccentricity
does not undergo a random walk, but instead tends to higher eccentricities when
the binary is hard but before gravitational radiation circularization is
important.  As discussed in section 4 of \cite{ASF06}, a residual eccentricity
will induce a difference in the phase evolution of the second harmonic compared
to a circular orbit, even if it is as small as 0.07, as \cite{ASF06} found. In
our case, if we use an eccentricity $e_{10^{-4}\,{\rm Hz}}=0.3$ in equation (4)
of \cite{ASF06}, we find that the accumulated phase shift $\Delta \Psi_{\rm e}
\ge 2\pi$ if observations cover a time of at least 

\begin{equation}
t_{\rm mrg} \sim 0.012\,\cdot(1+z)^{2}~{\rm yr}
\label{eq.Tmerg_z}
\end{equation}
\noindent
before merger, where $z$ is the redshift. If we set $z = 1$, then we have to
cover a time $t_{\rm mrg} = 17$ days before merger. This means that if
we are able to observe the system during that period of time before the final
coalescence, we will recover enough information to determine that the
orbit is not circular. On the other hand, if we use a residual 
eccentricity of 0.07, as in \cite{ASF06}, we would need 3-4 years of observation
for a $300+300\,M_{\odot}$ binary before merger.

In conclusion, if young massive clusters form binary IMBHs then they will be
strong and moderately eccentric LISA sources that could serve as unique
signposts of clustered star formation.   The non-zero residual eccentricity has
an impact on the detection of such sources, since it is generally assumed that
an equal-mass massive binary will have a zero eccentricity when entering the
LISA band. Our results show that in our scenario $e$ is non-negligible for
certain cases -though for some other models it is very low but detectable-;
notably, case $C$ and $F$, which are the only models in which we have a mass
fraction and thus, they are the more realistic ones. The process of formation
must of course be studied carefully from the standpoints of stellar dynamics
and merger product evolution, but if binary IMBHs can form then their mergers
are promising sources for future LISA detections.

\acknowledgments
We thank Vanessa Lauburg for valuable discussions.
We are also indebted to Yuri Levin,
Ed Porter, Matt Benacquista, Jonathan Gair and Stas Babak for enlightening
conversations.  The authors are grateful to the Aspen Center for Physics
for its hospitality during part of this project.
PAS work was partially supported by the MEC (Ministerio de
Educaci{\'o}n y Ciencia) at the Institut de Ci{\`e}ncies de l'Espai (IEEC/CSIC)
and the DLR (Deutsches Zentrum f\"ur Luft- und Raumfahrt) at the Max-Planck
Institut f\"ur Gravitationsphysik (Albert Einstein-Institut, AEI). MCM
appreciates support from NASA under ATFP grant NNX08AH29G. The
work of MF is funded through the PPARC rolling grant at the Institute of
Astronomy (IoA) in Cambridge. The simulations have been performed at the GRAPE
cluster {\sc Tuffstein} of the AEI.  


\label{lastpage}

\end{document}